\documentstyle[manuscript,aps]{revtex}

\input epsf

\def\tr{{\rm tr}\,}
\newcommand{\beq}{\begin{equation}}
\newcommand{\eeq}[1]{\label{#1} \end{equation}}
\newcommand{\beqar}{\begin{eqnarray}}
\newcommand{\eeqar}[1]{\label{#1} \end{eqnarray}}

\newcommand{\sn}{{\rm sn}}
\newcommand{\cn}{{\rm cn}}
\newcommand{\dn}{{\rm dn}}
\newcommand{\diag}{{\rm diag}}

\begin{document}
\draft

\preprint{CU-TP-856, {\tt hep-th}/9709080}

\setcounter{page}{0}

\title{\Large\bf Two Massive and One Massless $Sp(4)$ Monopoles} 

\author{\large\it
Kimyeong Lee\footnote{electronic mail: klee@phys.columbia.edu}
and Changhai Lu\footnote{electronic mail: chlu@cuphy3.phys.columbia.edu}}

\address{Physics Department, Columbia University, New York, NY, 10027}
\date{\today}

\maketitle

\begin{abstract}
Starting from Nahm's equations, we explore BPS magnetic monopoles in
the Yang-Mills Higgs theory of gauge group $Sp(4)$ which is broken to
$SU(2)\times U(1)$. A family of BPS field configurations with purely
Abelian magnetic charge describe two identical massive monopoles and
one massless monopole. We construct the field configurations with
axial symmetry by employing the ADHMN construction, and find the
explicit expression of the metrics for the 12-dimensional moduli space
of Nahm data and its submanifolds.
\end{abstract}


\setcounter{footnote}{0}

\section{Introduction}

In this paper we consider the Yang-Mills Higgs theory whose gauge
symmetry $Sp(4)$ is broken to $SU(2)\times U(1)$. We investigate a
family of purely Abelian configurations which describe two identical
massive and one massless monopoles. We approach the problem by solving
Nahm's equations under proper boundary and compatibility conditions.
By using the Atiyah-Hitchin-Drinfeld-Mannin-Nahm (ADHMN)
construction~\cite{adhm,nahm}, we construct the field configurations
in spherically and axially symmetric cases.  We then calculate the
metrics of the 12-dimensional moduli space $M^{12}$ of Nahm data and
its submanifolds.  Generally it is expected that the moduli space of
Nahm data is isometric to the moduli space of the corresponding
monopole configurations. We examine the metric of the moduli space in
detail and show that it behaves consistently with what is expected
from the dynamics of monopoles.

Recently magnetic monopoles have again become a focus of attention as
they play a crucial role in study of electromagnetic duality in the
supersymmetric Yang-Mills theories. The relevant magnetic monopole
solutions are of the Bogomol'nyi-Prasad-Sommerfield (BPS) type such
that the static interaction between magnetic monopoles
vanishes~\cite{bps}. The gauge inequivalent field configurations of
the BPS monopole solutions are characterized by the moduli parameters
associated with the zero modes of the solutions.  The metric of the
moduli space determines the low energy dynamics of
monopoles~\cite{manton1}. The electromagnetic duality has been
explored by studying quantum mechanics on the moduli space of the BPS
monopoles.

When the gauge group is not maximally broken so that there is an
unbroken non-Abelian gauge symmetry, the moduli space dynamics becomes
more subtle because of the global color
problem~\cite{color}. Nevertheless it has been known that the moduli
space is well defined when the total magnetic charge is purely
Abelian~\cite{weinberg1}.  Recently some of such moduli spaces have
been studied by starting from the maximal symmetry breaking case and
restoring the broken symmetry partially~\cite{kleen}. {}From this
point of view some magnetic monopoles become massless, forming a
non-Abelian cloud surrounding remaining massive monopoles. The global
part of unbroken gauge symmetry becomes the isometry of the moduli
space.  The meaning of the moduli space coordinates of massless
monopoles changes from their positions and phases to the gauge
invariant structure parameters for the cloud and gauge orbit
parameters.  With a inequivalent symmetry breaking $Sp(4)\rightarrow
SU(2)\times U(1)$, an Abelian combination is made of one masssive and
one massless monopoles. This simple case where the field configuration
and the moduli space metric are completely known  was studied in detail to
learn about the non-Abelian cloud~\cite{kleen,ejwso5}.

The next nontrivial purely Abelian configurations beyond this simple
model are made of two massive and one massless monopoles. Two massive
monopole can be distinguished as in the example where
$SU(4)\rightarrow U(1)\times SU(2)\times U(1)$. In that case, the
so-called Taubian-Calabi metric for the moduli
space~\cite{kleen,rychenkova,rocek} is be obtained from the massless
limit of that of the maximally broken case~\cite{klee2}. Two massive
monopoles are identical in the cases where $SU(3), Sp(4), G_2
\rightarrow SU(2)\times U(1)$.  (See Table~I and II of
Ref.~\cite{kleen}.)  Sometime ago the moduli space in the case
$SU(3)\rightarrow SU(2) \times U(1)$ has been found by Dancer by
exploring the moduli space of Nahm data~\cite{dancer1,dancer2}.

Our approach is similar to Dancer's. We use the embedding
procedure to construct $Sp(4)$ configurations from $SU(4)$
configurations. Some of the field configurations are simpler than
Dancer's.  Our spherical symmetric solution is just an embedding of
the $SU(2)$ solution. A class of our axially symmetric solutions can
be obtained from a linear superposition of configurations for two
noninteracting monopoles. Our work provides a further illustration of
the role of massless monopoles.

Another motivation for studying the moduli space of configurations
involving massless monopoles is that it may lead us to some new
insight about mesons and baryons in quenched QCD. Even in quenched
QCD, nondynamical external quarks are expected to be confined and form
mesons and baryons. Suppose that quenched QCD has been
supersymmetrized to $N=4$ so that there is no confinement. (Here we
imagine that all supersymmetric partners are very massive initially and
then become light.)  If the coupling constant is still strong, the
resulting configurations of mesons and baryons cannot be described by
Coulomb potentials as the nonlinear gauge interaction is not
negligible. The non-Abelian gauge field should somehow form a cloud
around external quarks, making the whole configuration to be a gauge
singlet, because of the continuity of the configuration with respect
to the mass parameter.  The shape of this cloud may remember 
confinement strings which connected the quarks. This can be regarded
as the limit where  confining string becomes tensionless.

If the electromagnetic duality holds even when the unbroken gauge
symmetry is partially non-Abelian~\cite{gno}, mesons and baryons can
have their magnetic dual, which are made of massive and massless
monopoles.  Indeed massive monopoles play the role of external quarks
and massless monopoles play that of non-Abelian cloud.  Thus Abelian
configurations made of two massive and one massless monopoles can be
regarded as dual mesons.  More interestingly, the moduli space of
three massive and three massless monopoles in the $SU(4)\rightarrow
SU(3) \times U(1)$ can be regarded as a magnetic dual of
baryons~\cite{kimyeong}. The structure of  the dual baryons may 
hint  a shape of the confinement strings connecting three quarks.

The plan of this work is as follows. In Sec.~II, we review the method
to find Nahm data for the classical group.  In Sec.~III, we study the
symmetry breaking pattern $Sp(4)\rightarrow SU(2)\times U(1)$, and
solve Nahm's equations with relevant boundary conditions.  In Sec.~IV,
we use the ADHMN method to construct the Higgs field configurations in
spherically and axially symmetric cases. This leads to a general
understanding of the parameter space in terms of the size of
non-Abelian cloud and the distance between massive monopoles.  In
Sec.~V, we find the explicit metrics of the moduli space and its
submanifolds.  In Sec.~VI, we conclude with some remarks.

\section{Nahm Data}

The Bogomol'nyi equations satisfied by BPS monopoles can be
written as self-dual Yang-Mills equations
\begin{equation}
F_{\mu\nu}=\frac{1}{2}\epsilon_{\mu\nu\rho\sigma}F_{\rho\sigma}
\label{selfdual}
\end{equation}
in $R^4$ with coordinates $x_1, x_2, x_3,x_4$. All the fields of BPS
monopoles here depend only on $x_1, x_2, x_3$. Instead if everything
depends only on the complementary variable $x_4=t$, then
Eq.~(\ref{selfdual}) leads to the so-called Nahm's equations,
\beq
\frac{dA_i}{dt}+[A_4, A_i]=\frac{1}{2}\epsilon_{ijk}[A_j, A_k],
\eeq{2}
where $i,j,k=1,2,3$.  The solutions of Nahm's equations satisfying
certain boundary condition that will be stated below are called Nahm
data. We can always perform a gauge transformation to eliminate $A_4$,
so sometimes $A_4$ is not included in Nahm's equations.  Nahm's
equations are much easier to solve than the original self-dual
Yang-Mills equations, since they are ordinary differential
equations. The relationship between Bogomol'nyi equations (depend on
three variables) and Nahm's equations has been thoroughly investigated
especially in $SU(2)$ gauge group case~\cite{nahm,hitchin}.  There is
 a kind of duality between $d$ and $4-d$ dimensional self-dual
theories~\cite{corrigan}.  It is also believed in general that the
moduli spaces of Nahm data and BPS monopoles are isometric to each
other, which has been proven in the $SU(2)$ case~\cite{nakajima}. The
idea is that Nahm's equations are regarded as an infinite dimensional
moment map and that the hyperk\"ahler quotient~\cite{hitchin2} of the
infinite dimensional flat space will lead to the natural hyperk\"ahler
metric for the moduli space of Nahm data~\cite{nakajima,atiyah}.

The original Nahm's method of $SU(2)$ monopoles has been generalized
into all types of classical groups~\cite{group}.  Let's start with the
$SU(N)$ case since all other groups can be  treated by embedding them into
$SU(N)$. Assuming the asymptotic Higgs field is
$\phi_{\infty}=\diag(\mu_1, \cdots, \mu_N)$ with $\mu_1<\cdots<\mu_N$
along a given direction, then the Nahm data for multi-monopoles
carrying charge $ (m_1, \cdots, m_{N-1})$ are defined as $N-1$ triples
$({}^lT_1, {}^lT_2, {}^lT_3) \; (l=1, \cdots, N-1)$ satisfying:
\begin{enumerate}
\item For each $l$, ${}^lT_i \; (i=1, 2, 3)$ are analytic $u(m_l)$-valued 
functions satisfying Nahm's equations in interval 
$(\mu_l, \mu_{l+1}), \; l=1, \cdots, N-1$.
\item The boundary conditions relating the Nahm data in two adjoint
intervals are the following:
\begin{enumerate}
\item If $m_{l}>m_{l-1}$, then there exist non-singular limit,
$\lim_{t\rightarrow \mu_l^{-}}{}^{l-1}T_i={}^{l-1}S_i$ and the
structure of  ${}^{l}T_i$ near $t=\mu_l$ is
\begin{equation}
\lim_{t\rightarrow \mu_l^+}{}^{l}T_i=\left(\begin{array}{cccc}
{}^{l-1}S_i & \ast \\
\ast & \frac{{}^{l}R_i}{t-\mu_l}
\end{array}\right),
\label{boundc}
\end{equation}
where ${}^{l}R_i$ form an ($m_{l}-m_{l-1}$)-dimensional
irreducible representation of $su(2)$ (unless $m_{l}-m_{l-1}=1$ in
which case ${}^{l}R_i/(t-\mu_l)$ has to be replaced by a non-singular
expression), and ``$\ast$'' refers to the elements that are not interested
in this paper.
\item If $m_{l}<m_{l-1}$, the roles of $(\mu_{l-1}, \mu_l)$ and 
$(\mu_{l}, \mu_{l+1})$ are reversed. 
\item If $m_{l}=m_{l-1}$, the condition is more complicated but fortunately
we are not going to confront this situation in this paper.
\end{enumerate}
\end{enumerate}

The way to embed the cases of $SO(N)$ and $Sp(N)$ into $SU(N)$ group
is described in Table~1.  These embedding procedures are obtained by
constraining the $SU(N)$ generators further.  The generators $T$ of
$Sp(N)$ satisfy the condition $T^T J+J T=0$ such that $JJ^*=-I$. The
generators $T$ of $SO(N)$ satisfy the condition $T^T K+KT=0$ such that
$KK^*=I$. The explicit forms of $J,K$ can be deduced from Table~1. 

\vspace{4mm}
\begin{center}
\begin{small}
\begin{tabular}{|c|c|c|c|}\hline
$G$ & $G$-charge & $\phi_{\infty}$ in $SU(N)$ &
$SU(N)$-charge \\ \hline 
$Sp(N)$ & $\rho_1, \cdots, \rho_n$ & 
$\mu_l=-\mu_{2n+1-l}$ & $m_l=m_{2n-l}=\rho_l$ \\ 
$N=2n$&& $l=1, \cdots, n$ & $l=1, \cdots, n$ \\ \hline
$SO(N)$ & $\rho_1, \cdots, \rho_{n-2}$ &
$\mu_l=-\mu_{2n+1-l}$ & $m_l=m_{2n-l}=\rho_l$ \\
$N=2n$&$\rho_+, \rho_-$ &  $l=1, \cdots, n$ & $l=1, \cdots, n-2$\\
&&&$m_{n-1}=m_{n+1}=\rho_+ +\rho_-$ \\
&&&$m_n=2\rho_+$ \\ \hline
$SO(N)$& $\rho_1, \cdots, \rho_n$ & $\mu_l=-\mu_{2n+2-l}$
& $m_l=m_{2n+1-l}=\rho_l$\\
$N=2n+1$&&$l=1, \cdots, n+1$ & $l=1, \cdots, n-1$\\
&&& $m_n=m_{n+1}=2\rho_n$ \\ \hline 
\end{tabular}
\end{small} \\
{\bf Table 1}: The embedding of $Sp(N),\, SO(N)$ in $SU(N)$. 
\end{center}

\vspace{8mm}

These embedding procedures enable us to get the $SO(N)$ and $Sp(N)$ 
Nahm data from the $SU(N)$ data with asymptotic Higgs field
$\phi_{\infty}=\diag(\mu_1,...,\mu_N)$ and the charge $\{m_l\}$. What is 
new is that we now have one more set of conditions connecting the Nahm data
between different intervals:

\begin{enumerate}
\item[3.] There exist matrices ${}^lC \; (l=1, \cdots, N-1)$ satisfying
\begin{equation}
{}^{N-l}T_i(-t)^T= ({}^lC)\, {}^lT_i(t)\, ({}^lC^{-1}),
\label{compatible}
\end{equation}
and compatibility conditions:
\begin{enumerate}
\item ${}^{N-l}C={}^lC^T, \hspace{1cm} {\rm for} \hspace{2mm} Sp(N)$
\item ${}^{N-l}C=-{}^lC^T. \hspace{1cm} {\rm for} \hspace{2mm} SO(N)$
\end{enumerate}
These compatibility conditions reflect the fact that we are
identifying certain $SU(N)$ monopoles to get $SO(N)$ and $Sp(N)$
monopoles.
\end{enumerate}

In the above discussions we have assumed $\mu_1<\cdots<\mu_n$, which
physically means the gauge symmetry is maximally broken. We can also
consider the cases with non-Abelian unbroken symmetry so that some
$\mu_l$'s are equal, geometrically this is the case when some of the
intervals shrink to zero length. The monopole mass is proportional to
the size of the corresponding interval and so the shrunken intervals
are corresponding to massless monopoles. All the procedures described
above remain unchanged even in this case.

\section{Nahm Data in the  {\mbox{$Sp(4)$}} Case}

The model we consider is the $Sp(4)$ Yang-Mills theory with a single
Higgs field in the adjoint representation and no potential.  The
vacuum expectation value of the Higgs field is nonzero and the gauge
symmetry is spontaneously broken to $SU(2)\times U(1)$.  The roots and
coroots of the $Sp(4)=SO(5)$ group is shown in Fig.~1. Note that in
our convention $\alpha^*=\alpha/|\alpha|^2=\alpha$.

\begin{center}
\leavevmode
\epsfxsize=2.0in
\epsfbox{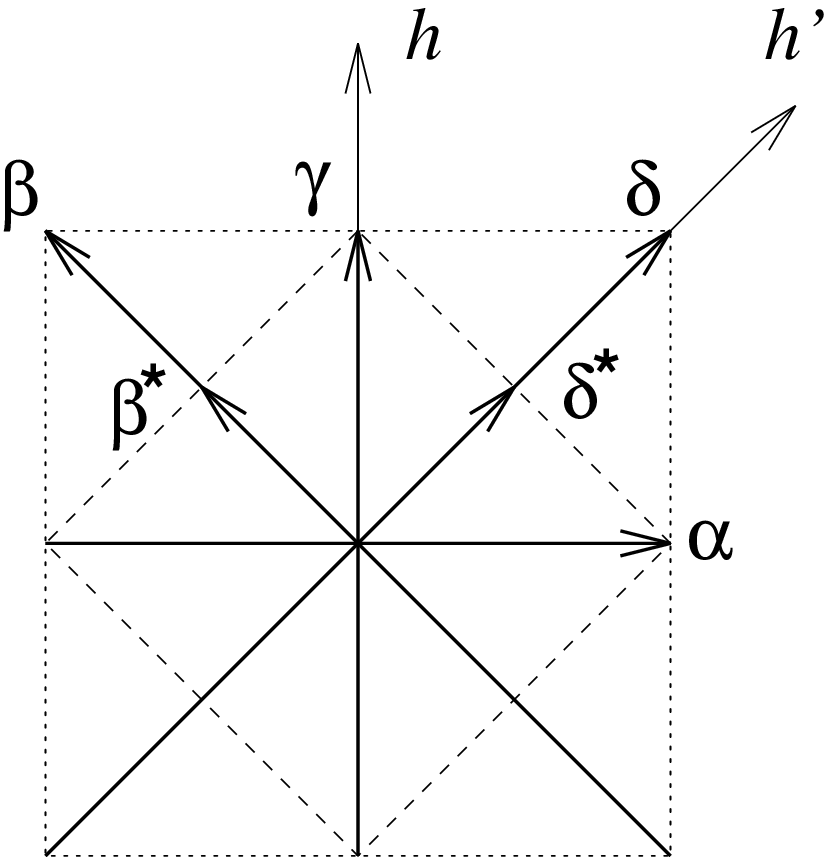}
\end{center}
\centerline{{\bf Figure 1}: The root diagram of $Sp(4)$}
\vspace{0.5cm}

In this paper we consider the symmetry breaking with
$\langle\Phi\rangle =h\cdot H $. The simple roots we choose for
convenience is $\beta,\alpha$ rather than $\delta,-\alpha$.  For any
root $\alpha$, there is a corresponding $SU(2)$ subalgebra,
\begin{eqnarray}
& & t^1(\alpha) = \frac{1}{\sqrt{2\alpha^2}}(E_\alpha+E_{-\alpha}),
\nonumber \\ 
& & t^2(\alpha)=\frac{-i}{\sqrt{2\alpha^2}}
(E_\alpha-E_{-\alpha}),\nonumber \\
& & t^3(\alpha)=\alpha^*\cdot H.
\end{eqnarray}
Using this $SU(2)$ algebra, we can embed the $SU(2)$ single monopole
solution along any root. Thus there is a spherically symmetric
monopole configuration for any root $\alpha$ such that $\alpha\cdot
h\neq 0$.  Since $\beta\cdot h>0$, the monopole with magnetic charge
$\beta^*$ is massive. (Here we are dropping the coupling constant
$4\pi/e$.)  On the other hand $\alpha^*\cdot h=0$ and so there is no
monopole solution corresponding to the root $\alpha$.  As argued in
the introduction, the zero mode counting can be done consistently only
for purely Abelian configurations.  In our case the simplest case has
the magnetic charge 
\begin{equation}
\gamma^*= 2\beta^*+\alpha^*,
\label{gamma}
\end{equation}
so that $\gamma^*\cdot \alpha=0$.  The moduli space of this
configuration is 12-dimensional and denoted by $ M^{12}$. As discussed
in Ref~\cite{kleen}, we imagine the $h$ as a limit where
$h\cdot\alpha$ is positive but becomes infinitesimal. We can regard
$\alpha^*$ monopoles as massless, and so the $\gamma^*$ monopole can
be thought as a composite of two identical massive $\beta^*$ monopoles
and one massless $\alpha^*$ monopole. Here we can see the internal
unbroken gauge group should be $SO(3)_g$ rather than $SU(2)$, because
all the generators of $Sp(4)$ transforms as vector or singlet
representations under the unbroken generators ${\bf t}(\alpha)$.

If we have chosen the Higgs expectation value as $h'$, the unbroken
$SU(2)$ would be associated with $\beta$. The Abelian configuration
could have the magnetic charge $\delta^*= \alpha^*+\beta^*$ such that
$\delta^*\cdot\beta = 0$. This configuration can be interpreted as a
composite of one massive $\alpha^*$ and one massless $\beta^*$
monopoles. The BPS field configuration and 8-dimensional moduli space
of this magnetic charge are known explicitly to be flat $R^4$. This is
the model which lead to many insights about non-Abelian
cloud~\cite{kleen}.

As discussed in Sec.~II, Nahm data for $Sp(4)$ can be studied by
embedding $Sp(4)$ in $SU(4)$. Thus the Higgs field can be written as
$4\times 4$ traceless Hermitian matrix.  As shown in Table 1, the
Higgs expectation value can be chosen to be $\langle \Phi \rangle =
\diag (-\mu_1,-\mu_2,\mu_2,\mu_1)$ with $\mu_1\ge \mu_2\ge 0$. Any
generator $T$ of the $Sp(4)$ subgroup should be traceless 
antihermitian and satisfy
\begin{equation}
TJ + JT^T= 0, 
\end{equation}
where the $Sp(4)$ invariant tensor $J$ is chosen to be
\begin{equation}
J = \left(\matrix{0&0&0&1 \cr
                  0&0&1&0 \cr
                  0&-1&0&0 \cr
                  -1&0&0&0 \cr} \right).
\label{jmatrix}
\end{equation}
This defines the $Sp(4)$ embedding in $SU(4)$ uniquely, which is also
consistent with Table~1.  A consistent choice of the Cartan subgroup
of $Sp(4)$ is $H_1={\rm diag}(-1,1,-1,1)/2$ and
$H_2=\diag(-1,-1,1,1)/2$. The two inequivalent symmetry breaking
patterns for $Sp(4)\rightarrow SU(2)\times U(1)$ in Fig.~1 correspond
to $h\cdot H =\diag(-1,-1,1,1)$ and $h'\cdot H
=(-1,0,0,1)=H_1+H_2$. Thus our case with $\mu_1=\mu_2=0$ corresponds
to the case where $SU(4) \rightarrow SU(2)\times U(1)\times SU(2)$.

{}From Table 1 in Sec.~II, we read that our configuration
(\ref{gamma}) in $Sp(4)$ has the $SU(4)$ magnetic charge $(1,2,1)$,
that is, two identical massive monopoles and two distinct massless
monopoles. This is exactly the configuration considered by
Houghton~\cite{houghton}, whose focus was on its hyperk\"ahler
quotient space.  If we have chosen the expectation value $h'$, the
simplest Abelian configurations have the magnetic charge $(1,1,1)$ in
$SU(4)$, that is, two distinct massive monopoles and one massless
monopole, whose moduli space metric has been found as the
Taubian-Calabi metric~\cite{kleen,rychenkova,rocek}.

According to the previous section, Nahm data $T_\mu(t)$ defined on the
interval $[-1,1]$ are anti-Hermitian two-by-two matrices and satisfy Nahm's
equations
\begin{equation}
\frac{dT_i}{dt} + [T_4,T_i] = \frac{1}{2}\epsilon_{ijk}[T_j,T_k],
\label{nahm2}
\end{equation}
and the compatibility condition
\begin{equation}
 T_\mu(-t)^T = C T_\mu(t)C^{-1}
\label{cond}
\end{equation}
with a symmetric matrix $C$.  The Nahm data should be analytic at the
end points $t=\pm 1$. The boundary and compatibility
conditions~(\ref{boundc}) and (\ref{compatible}) satisfied by the above
Nahm data become
\begin{equation}
(T_\mu(-1))_{11} = (T_\mu(1))_{22}
\label{bound}
\end{equation}
A detailed understanding of the boundary condition will be needed in
the case where the massless monopole becomes massive.

The space of Nahm data has the following symmetries:\hfill\break
\noindent 1. Local gauge transformations ${\cal G}=\{g(t)\in U(2)\} $
whose transformations are 
\begin{eqnarray}
& &T_4\rightarrow gT_4g^{-1}-\frac{dg}{dt}g^{-1}, \nonumber \\
& & T_i \rightarrow gT_i g^{-1}.
\label{gauge1}
\end{eqnarray}
They should be consistent with the conditions~(\ref{cond}) and
(\ref{bound}).  Its subgroup is $ {\cal G}_0 = \{g\in {\cal G}:
g(-1)=g(1)=1\}$. \hfill\break
\noindent 2. Spatial translation group $R^3$ with three parameters
$\lambda_i$:
\begin{eqnarray}
& &   T_4\rightarrow T_4 \nonumber \\
& & T_j\rightarrow T_j -i\lambda_j I. 
\end{eqnarray}
\noindent 3. Spatial rotation group $Spin(3)=\{ a_{ij}\in SO(3)\}$:
\begin{equation}
T_i\rightarrow \sum_j a_{ij}T_j.
\label{rot}
\end{equation}
Notice that Eq.(\ref{rot}) is a pure rotation as there is no residues
to be fixed  at $t=\pm 1$.  (This indicates that the rotational group is
$SO(3)$ rather than $SU(2)$.)

To solve Nahm's equations together with  the compatibility condition, we use
the spatial translation to make $T_\mu$ traceless.  This traceless Nahm
data is called centered and describes the monopole configuration in
the center of mass frame.  We can also choose the gauge $T_4=0$.
Furthermore we use the spatial rotation to set the $t$-independent
$\tr(T_1T_2)$, $\tr(T_1T_3)$ and $\tr(T_2T_3)$ to be zero. After a
spatial rotation, we get that for each $j=1,2,3$;
\begin{equation}
T_j=\frac{1}{2}f_j\tau_j,
\label{data0}
\end{equation}
where quaternions $\tau_j$ are chosen to be 
\begin{equation}
\tau_1=\left( \begin{array}{cc}
                 i & 0 \\
                 0 & -i 
              \end{array}\right),\,\,
\tau_2=\left( \begin{array}{cc}
                 0 & 1 \\
                 -1 & 0 
              \end{array}\right),\,\,
\tau_3=\left( \begin{array}{cc}
                 0 & i \\
                 i & 0 
              \end{array}\right),
\end{equation}
satisfying $\tau_1\tau_2 =\tau_3$, etc.
Then, Nahm's equations 
become the well-known Euler top equations:
\begin{eqnarray}
& & \dot{f_1}=f_2 f_3,\nonumber\\
& &  \dot{f_2}=f_3 f_1, \nonumber\\
& & \dot{f_3}=f_1 f_2.
\end{eqnarray}
We note that $f_1^2-f_2^2$ and $f_2^2-f_3^2$ are independent of
$t$. Hence let us consider the case $f_1^2\le f_2^2\le f_3^2$. Then
the solution to this set of equations is known in terms of Jacobi
elliptic functions as
\begin{eqnarray}
f_1&=&-\frac{D\cn_k[D(t-t_0)]}{\sn_k[D(t-t_0)]}, \nonumber \\
f_2&=&-\frac{D\dn_k[D(t-t_0)]}{\sn_k[D(t-t_0)]},\nonumber\\
f_3&=&-\frac{D}{\sn_k[D(t-t_0)]},
\label{data1}
\end{eqnarray}
where $k\in [0,1]$ is the elliptic modulus and $D,t_0$ are arbitrary.
We can change the sign of any two of $f_1, f_2$ and $f_3$ by $180$
degree rotations.

On the other hand, the compatibility condition (\ref{cond}) becomes
that for every $j$,
\begin{equation}
f_j(-t)\tau_j^T = f_j(t) C\tau_j C^{-1}
\end{equation}
with a symmetric matrix $C$. The boundary condition (\ref{bound})
becomes $f_1(-1)=-f_1(1)$.  Among linear combinations of $\tau_1$ and
$\tau_3$, the right choice for $C$ with Nahm data (\ref{data1}) is
\begin{equation}
C=\tau_3. 
\end{equation}
This implies that $f_1$ is an odd function and $f_2,f_3$ are even
functions.\footnote{If $f_1^2$ is not chosen to be smallest, it would
contradict with  the boundary condition (\ref{boundc}).}  This fixes the
parameter $t_0$ to satisfy $\cn_k(Dt_0)=0$. Then our solution for
Nahm's equation is as follows:
\begin{eqnarray}
& & f_1=  D\sqrt{1-k^2}\frac{\sn_k(Dt)}{\cn_k(Dt)} ,\nonumber \\\
& & f_2= -D\sqrt{1-k^2}\frac{1}{\cn_k(Dt)},\nonumber\\ 
& & f_3= -D\frac{\dn_k(Dt)}{\cn_k(Dt)}.  
\label{data2}
\end{eqnarray}
This Nahm data is regular for $t\in [-1,1]$.  The analyticity of the
data requires that $0\le k\le 1$ and $0\le D\le K(k)$ with $4K(k)$
being the period of $\sn_k$. $K$ is also the first complete elliptic
integral $K(k)=\int_0^{\pi/2} d\theta ( 1-k^2\sin^2\theta)^{-1/2}$.
Eqs.~(\ref{data0}) and (\ref{data2}) are the Nahm data we are looking
for.  (Actually they are the Nahm data on a representative point of
the  $SO(3)\times SO(3)$ orbit.)  Sometimes we will simply call
Eq.~(\ref{data2}) as Nahm data. There are eight equivalent copies of
above Nahm data: we can exchange $f_2$ and $f_3$, and any two of
$f_1,f_2$ and $f_3$ can change their sign. The allowed local gauge
transformations of Eq.~(\ref{gauge1}) are made of $g(t)$ such that
\begin{equation}
g(t)= e^{ \epsilon_j(t)\tau_j/2}
\label{gaugetr}
\end{equation}
with even $\epsilon_1$ and odd $\epsilon_2, \epsilon_3$ functions.
This will be crucial in showing the spherically symmetric Nahm data is
not invariant under global gauge transformations due to $\epsilon_2,
\epsilon_3$.

The moduli space $M^{12}$ of uncentered three monopoles is the space
of gauge inequivalent Nahm data with the gauge action ${\cal G}_0$.
Since the center $U(1)$ of $U(2)$ is tri-holomorphic, we can perform a
hyperk\"ahler quotient with the momentum map $\mu = -i(\tr T_1,\tr
T_2,\tr T_3)$. This gives the eight-dimensional relative moduli space
$M^8$ of the centered Nahm data.  Further quotient of this manifold by
the internal gauge symmetry $SU(2)$ leads to the five dimensional
manifold $N^5=M^{8}/SU(2)$. The homeomorphic coordinates for $N^5$ are
given in terms of gauge-invariant $t$-independent
quantities~\cite{dancer1},
\begin{eqnarray} 
 \lambda_1&=&\langle T_1, T_1
\rangle-\langle T_2, T_2 \rangle, \nonumber\\
 \lambda_2&=&\langle T_1, T_1
\rangle-\langle T_3, T_3 \rangle, \nonumber \\ 
\lambda_3&=&\langle T_1, T_2
\rangle, \nonumber \\ 
\lambda_4&=&\langle T_1, T_3 \rangle, \nonumber\\
 \lambda_5&=&\langle T_2, T_3 \rangle,
\end{eqnarray}
where 
\begin{equation}
<T,T'> = -\int_{-1}^1 dt\, \tr (TT').
\end{equation}
They form a real traceless $3\times 3$ matrix and realize a
5-dimensional representation of $SO(3)$. The data (\ref{data2})
leads to the coordinates,
\begin{eqnarray}
& & \lambda_1 = -(1-k^2)D^2, \nonumber \\
& & \lambda_2 = -D^2,
\label{lambda12}
\end{eqnarray}
and $\lambda_3=\lambda_4=\lambda_5=0$, which is invariant under the
180 degree rotations around three cartesian axes. Thus this data has
$Z_2\times Z_2$ isotropy group.  $N^5$ is a five dimensional manifold
homeomorphic to $R^5$ and admits a non-free rotational $SO(3)$ action.
Further quotient of this manifold by the spatial rotation group $SO(3)$
leads to a two dimensional surface $N^5/SO(3)$, whose eight copies, as
we will see in Sec.~V, make a geodesic complete manifold ${\cal Y}^2$.
There are also two-dimensional surfaces of revolution, which describe
axially symmetric configurations.

Since the gauge group $SU(2)$ is tri-holomorphic, there is another
hyperk\"ahler quotient of $M^8$. We choose a $U(1)$ subgroup which
fixes $\tau_1$. The corresponding moment map is
\begin{equation}
\mu =  (\tr (T_1(1)\tau_1),
\tr(T_2(1)\tau_1),\tr(T_3(1)\tau_1)).
\label{moment}
\end{equation}
The hyperk\"ahler quotent space $M^4({\mbox{\boldmath $\zeta$}})
=\mu^{-1}({\mbox{\boldmath $\zeta$}})/U(1)$ is a four-dimensional
hyperk\"ahler space.  The rotational transformation
$Spin(3)=\{a_{ij}\}$ generates a homeomorphic mapping $M^4(\zeta_i)$
to $M^4(a_{ij}\zeta_j)$. We will see later that this family
interpolates the flat space $M^4(0)=R^3\times S^1$ to the
Atiyah-Hitchin space $M^4(\infty)$.  This family can be regarded as
deformations of the Atiyah-Hitchin space. Since any hyperk\"ahler
space in four dimensions is self-dual and so Ricci flat,
$M^4(\zeta,0,0)$ can be regarded as one-parameter family of
gravitational instantons.

\section{The ADHMN Construction}

Given Nahm data, we can define the differential operator
\beq 
\Lambda^\dagger({\bf x})=i\frac{d}{dt} -\sum_{i=1}^3 (iT_j
+x_j)\otimes e_j ,
\eeq{18} 
where $e_j\; (j=1, 2, 3)$ are quaternion units. The dimension of the
kernel of $\Lambda^{\dagger}$ depends on the boundary conditions
involved in defining Nahm data $T_i$. For our case it turns out to
be four. The basis of ${\rm Ker}\,\Lambda^\dagger$ consists of four
orthonormal four-component vectors ${\bf v}_\mu, \mu=1,...4$ with the
inner products $<{\bf v}_\mu,{\bf v}_\nu>=\int_{-1}^1 dt\, {\bf
v}^\dagger_\mu \cdot {\bf v}_\nu =\delta_{\mu\nu}$.  In terms of the
$4\times 4 $ matrix $V=({\bf v}_1,{\bf v}_2,{\bf v}_3,{\bf v}_4)$, the
ADHMN construction of monopole solutions in $R^3$ goes as follows:
the $4\times 4$ Hermitian matrix-valued fields
\begin{eqnarray}
& &  \Phi =\int_{-1}^{1}dt \, t
V^{\dagger} V, \label{Higgs}\\
& &  A_j= i\int_{-1}^{1}dt\, V^\dagger
\frac{\partial V}{\partial x_j}, \label{vector}
\end{eqnarray}
form a BPS monopole field configuration. It is really a configuration
in $SU(4)$ gauge theory and may need a further gauge transformation in
$SU(4)$ to be expressed as a proper $Sp(4)$ configuration.

We express a single four vector as ${\bf v}=(w_1, w_2, w_3,
w_4)^T$. With the usual convention of quaternion units (namely
$e_1=\tau_1,\; e_2=\tau_2,\; e_3=\tau_3$), the equation
$\Lambda^{\dagger}{\bf v}=0$ can be written as
\begin{eqnarray}
& &
\dot{w_1}-x_1w_1 -(x_3-ix_2)w_3+\frac{1}{2}f_1w_1+\frac{1}{2}(f_3-f_2)w_4=0,
\nonumber \\ & &
\dot{w_2}-x_1w_2-(x_3-ix_2)w_4-\frac{1}{2}f_1w_2+\frac{1}{2}(f_2+f_3)w_3=0,
\nonumber \\ & &
\dot{w_3}+x_1w_3-(x_3+ix_2)w_1-\frac{1}{2}f_1w_3+\frac{1}{2}(f_2+f_3)w_2=0,
\nonumber \\ & &
\dot{w_4}+x_1w_4-(x_3+ix_2)w_2+\frac{1}{2}f_1w_4+\frac{1}{2}(f_3-f_2)w_1=0.
\label{kernel}
\end{eqnarray}
It is hard to obtain  general solution of the above equations. In
this section, we would like to work out several special cases in order
to check whether the ADHMN construction leads to the sensible
result. This exercise also yields a general understanding of the physical
meaning of parameters $k$ and $D$ appearing in Nahm data.

The first case we consider  is the spherically symmetric solution with
$D=0$ and so
\beq
f_1=f_2=f_3=0.
\eeq{20}
Clearly this Nahm data is invariant under the spatial SO(3) rotation
(\ref{rot}).  One may wonder whether this Nahm data is invariant under
global gauge transformations (\ref{gauge1}). Clearly this data $T_i=0$
is invariant under the global $SO(3)$ gauge rotation
(\ref{gauge1}). However, the initial $T_4=0$ is not necessarily
invariant. The reason is that the gauge parameters $\epsilon_2,
\epsilon_2$ of Eq.~(\ref{gaugetr}) are odd functions and so their time
derivative is nonzero for nontrivial transformations. But $\epsilon_1$
is even and so can be constant, leaving $T_0$ invariant. Thus, one
expects a $S^2$ gauge orbit space, which leaves the spherically
symmetric solution.  This two sphere will also appear in the metric of
the moduli space in the next section. (In Dancer's case, the
spherically symmetric solution is not invariant for all three
generators of $SU(2)$ gauge rotation.)

The kernel equations (\ref{kernel})  can be easily solved for the
spherically symmetric solution and give rise to the Higgs field 
\begin{equation}
\Phi= 2{\rm H}(2r)\,\hat{\bf r}\cdot{\bf t}(\gamma),
\end{equation}
where $ r=\sqrt{x_ix_i}$, $ \hat{r}_i=x_i/r$ and ${\rm H}(r)=\coth
(r)-1/r$ is the famous single monopole function.  This is the well
known single monopole solution with $\Phi_{\infty}\propto H_2$ along
$x_1$ direction.  This configuration is the $SU(2)$ embedded solution
along the composite root $\gamma$.  The energy density is maximized at
the center. We just argued that the corresponding Nahm data is not
invariant under some of the global gauge transformations. To understand
this in terms of the field configuration, we deduce form the root
diagram in Fig.~1 that the generators $t^i(\gamma)$ commute with
$t^3(\alpha)$ but not with $t^{1,2}(\alpha)$. Thus the spherically
symmetric field configuration is not invariant under two of
$t^i(\alpha)$, argued before.

We now turn to the next simplest case, the axially symmetric case.
Similar as in Dancer's case, we have two types of axially
symmetric cases.  The hyperbolic case appears when $k=1$ and $0\le
D<\infty$ so that
\begin{equation}
f_1=f_2=0 ,\,\,f_3=D.
\label{hyper1}
\end{equation}
This Nahm data is invariant under rotation around the
$x_3$-axis. Although no hyperbolic function is involved here, we have
used the same terminology as used as in Ref.~\cite{dancer1}, because
of a similarity in the qualitative behavior.  The trigonometric case
appears when $k=0$, so that
\begin{equation}
f_1= D\tan (Dt),\,\, f_2=f_3=-D\sec(Dt) 
\label{trigo}
\end{equation}
with $0\leq D<\frac{\pi}{2}$.  This data is invariant under the
rotation around the  $x_1$-axis.

Our hyperbolic case is much simpler than the corresponding case
considered in Dancer's.  After solving Eq.~(\ref{kernel}), we use
Eq.~(\ref{Higgs}) and a gauge transformation to obtain the Higgs
configuration,
\begin{equation}
\Phi=2H(2r_+)\hat{{\bf r}}_+\cdot {\bf t}(\beta)+2H(2r_-)\hat{{\bf
r}}_-\cdot {\bf t}(\delta),
\label{hyp}
\end{equation}
where ${\bf r}_\pm=(x_1,x_2,x_3\pm D/2)$.  We recognize that this
configuration describes $\beta^*$ and $\delta^*$ monopoles located at
the $x_3$ axis with $x_3=-D/2$ and $x_3=D/2$, respectively. Since
$[t^i(\beta),t^j(\delta)]=0$, there is no interaction between these
two monopoles, and the field configuration (\ref{hyp}) is just a
superposition of two corresponding configurations. In Dancer's
hyperbolic case, two massive monopoles  are interacting. 

The above hyperbolic configuration is not invariant under global gauge
rotations of ${\bf t}(\alpha)$ as it does not commute with ${\bf
t}(\beta)$ and ${\bf t}(\delta)$.  Among the dyonic excitations, there
is a  simple one which is  just a superposition of $\beta$ dyon and
$\delta$ dyon. Once the magnitudes of their electric charges are not
identical, their relative charge is nonzero. This corresponds to the
excitation due to the $t^3(\alpha)$ rotation. Clearly this
configuration would preserve the axial symmetry. In the next section,
the motion which changes $D$ and this relative charge will be
described by a flat two dimensional surface of revolution. Especially
the configuration with relative electric charge is spherically
symmetric when $D=0$, which is consistent with the fact that the
spherically symmetric solution is not invariant under the global gauge
rotation.

On the other hand our trigonometric solution (\ref{trigo}) is more
complicated.  Equation~(\ref{kernel}) at $(z, 0, 0)$ becomes
\begin{eqnarray}
& & \dot{w_1}-z w_1+\frac{1}{2}D\tan (Dt) w_1=0,
\label{phi222}\\
& & \dot{w_2}-z w_2-\frac{1}{2}D\tan (Dt) w_2-D\sec (Dt) w_3=0,
\label{phi11} \\
& & \dot{w_3}+z w_3-\frac{1}{2}D\tan (Dt) w_3-D\sec (Dt) w_2=0, 
\label{phi44}\\
& & \dot{w_4}+z w_4+\frac{1}{2}D\tan (Dt) w_4=0.
\label{phi14}
\end{eqnarray}
Notice that the first and fourth equations are not coupled with 
anything else while the second and third equations are only coupled
among themselves. Thus,   after a $SU(4)$ gauge
transformation  the Higgs field  has the form
\begin{equation}
\Phi= \left( \begin{array}{cccc}
\ast & 0 & 0 & \ast \\ 0 & \ast & 0 & 0 \\ 
0 & 0 & \ast & 0 \\ \ast & 0 & 0 & \ast \end{array}\right),
\label{phia}
\end{equation}
where $*$ indicates nonvanishing entry.  Since $\Phi^T J +J \Phi = 0$
with $J$ in Eq.~(\ref{jmatrix}), we get $\Phi_{33} = -\Phi_{22}$ and
$\Phi_{44}=-\Phi_{11}$. {}From Eq.~(\ref{phi222}), we can easily obtain
\begin{equation}
\Phi_{22}=-\frac{f(z)-f(-z)}{g(z)+g(-z)},
\label{phi22}
\end{equation}
where
\begin{eqnarray}
& & f(z)=e^{2z}\left\{[(2z+1)D^2+4z^2(2z-1)]\cos D +D[D^2+4z(z-1)]\sin
D\right\}, \\
& & g(z)=e^{2z}(D^2+4z^2)(2z\cos D+D\sin D).
\end{eqnarray}
We are not going to pursue the details for the corner $2\times 2$
matrix part of $\Phi$, which describes the non-Abelian part.  Like in
the case of Ref.~\cite{dancer1}, we have a reason to believe that the
trigonometric data corresponds to the situation when the energy
density is maximized on a ring around the axis of symmetry, even
though we have not done the numerical computation to check this. When
$D=0$, the configuration is spherically symmetric. When $D\rightarrow
\pi/2$, we will see in a moment that our result approaches the
Atiyah-Hitchin case. That case, when axially symmetric, has the
ring-like energy distribution. Thus  symmetry and continuity imply
the ring-like energy distribution for the trigonometric case.

At the limit $D\rightarrow \pi/2$,  Eq.~(\ref{phi22}) becomes
\begin{equation}
\Phi_{22}=-\left[\tanh(2z)-\frac{z}{z^2+\left(\frac{\pi}{4}\right)^2}\right],
\end{equation}
which is exactly the result of two $SU(2)$ monopoles~\cite{ward}.
Meanwhile Eqs.~(\ref{phi11}) and (\ref{phi44}) lead to $\Phi_{11}= -
\Phi_{44}=-1$ and $\Phi_{14}=\Phi_{41}=0$ at $D=\pi/2$. Thus the Higgs
field (\ref{phia}) along the symmetric axis becomes the Higgs field
for charge two $SU(2)$ monopole configuration.

As a general verification of the suggestion made above, let us check
whether the three monopole case degenerates into the $SU(2)$ result
when $k=0, D=\pi/2$, or more generally when $D\rightarrow K(k)$.
{}From Nahm data (\ref{data2}), we get
\begin{equation}
f_1,f_2,f_3 \approx -\frac{1}{1+t},
\end{equation}
near $t=-1$ and 
\begin{equation}
-f_1,f_2,f_3 \approx  -\frac{1}{1-t}
\end{equation}
near $t=1$.  These are exactly the boundary conditions satisfied by
Nahm data for two identical monopoles in the $SU(2)$
case~\cite{nahm,hitchin}. The removal of massless monopole to spatial
infinity leaves two identical monopoles and so corresponds to the
$SU(2)$ limit $D\rightarrow K(k)$.

Now let us  try to find  the physical meaning of two parameters $k$
and $D$ from the above mentioned solutions.  As we know, there is no
spherically symmetric solution for two identical monopoles in $SU(2)$
case. Thus,  non-Abelian cloud must play a crucial role in our
spherically symmetric solution with $D=0$.  Our spherically symmetric
solution is the embedding of the $SU(2)$ solution along the $\gamma$
root.  Thus the position of the massless monopole is at origin,
implying the minimum  cloud size. In the opposite limit where
$D\rightarrow K(k)$, Nahm data becomes that of the $SU(2)$
case, implying  non-Abelian cloud has been removed to spatial
infinity. Indeed it seems that non-Abelian cloud size increases
monotonically with $D$ for a given $k$. (We should add that the exact
nature of non-Abelian cloud is still uncertain when the cloud size is
comparable with distance between the massive monopoles.)

In the trigonometric case with $k=0$, the spherically symmetric
solution changes to the ring shape, as we take out massless monopole
to the spatial infinity by increasing $D$ from zero to $\pi/2$. 
In the hyperbolic case with $k=1$, the cloud is always in its minimum
size, so that the field configuration becomes effectively that of two
noninteracting $\beta^*$ and $\delta^*$ monopoles.

{}From these analysis we can also see the meaning of $k$. At least in
the Atiyah-Hitchin case we know that small $k$ corresponds to small
separation and $k\approx 1$ corresponds to wide separation, in which
case the distance goes roughly like $r\approx K(k)$. Even for
other $D$ we could expect $k$ to be related to the distance between
two massive monopoles in a qualitatively similar way as in
Atiyah-Hitchin case.  Of course when non-Abelian cloud has finite
size, the separation between massive monopoles cannot go beyond the
scope of cloud so $k=1$ would no longer represent the infinite
distance. Figure~2 shows the $k-D$ space. The spherically symmetric
case corresponds to $D=0$ and the trigonometric case does to the line
$k=0$ and $0<D<\pi/2$. The hyperbolic case corresponds to $k=1$ and
the Atiyah-Hitchin case does to the curve $D=K(k)$.

\begin{center}
\leavevmode
\epsfxsize=2.0in
\epsfbox{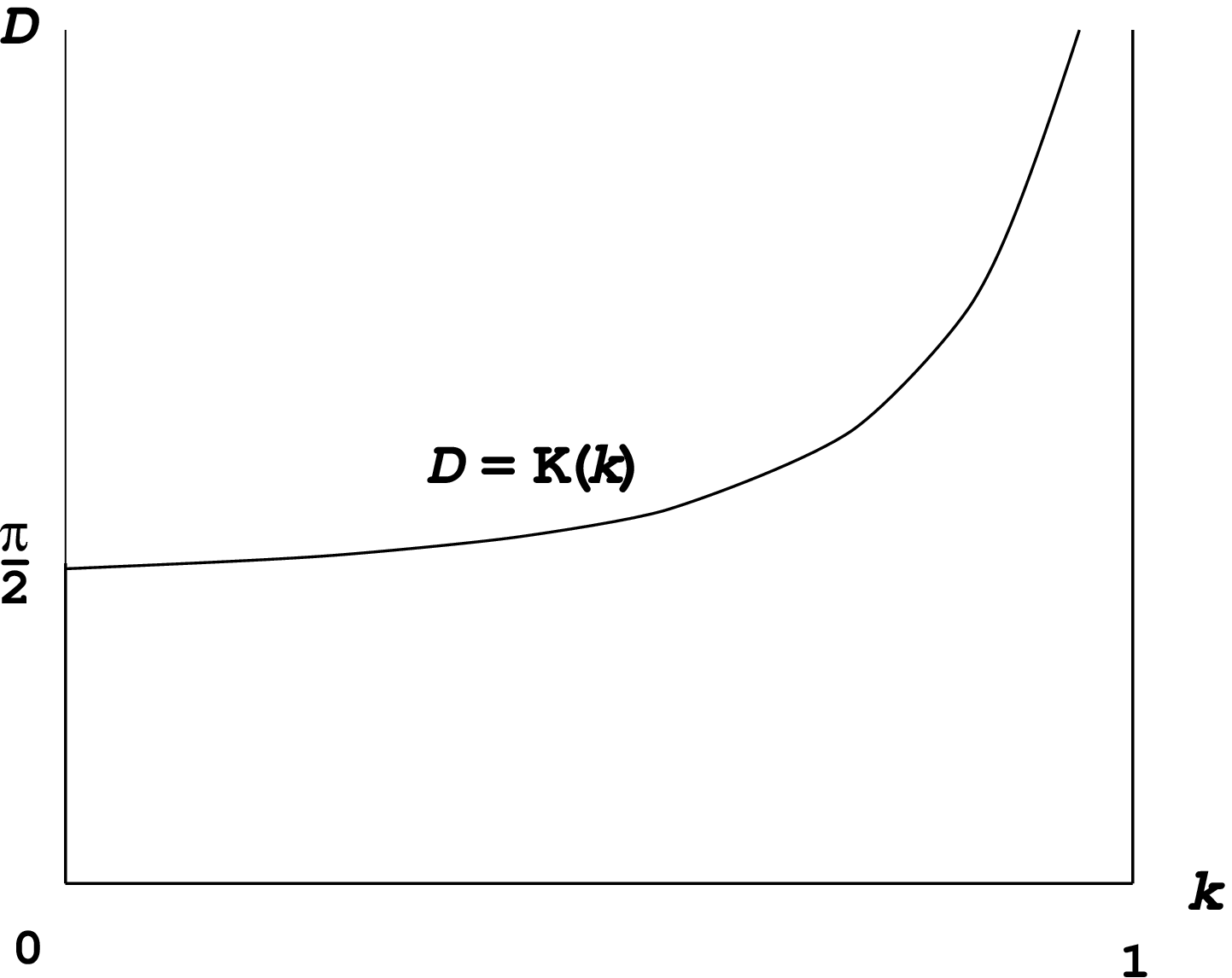}
\end{center}
\centerline{{\bf Figure 2}:  The $k-D$ space.}

\section{The Moduli Space Metric}

Now let us turn our attention to the metric of the moduli space. By
using centered Nahm data, we work in the center of mass frame of
monopoles. The relative moduli space $M^8$ of Nahm data should
isometrically correspond to the relative moduli space of the monopole
dynamics.  The metric for
the center of mass motion is flat, and we expect that 
\begin{equation}
M^{12} = R^3\times \frac{S^1\times M^8}{\Delta}
\end{equation}
where $\Delta$ is a discrete subgroup, about which we will not concern
in here.  Our work of finding the moduli space metric is greatly
facilitated by the works done by Dancer~\cite{dancer1} and
Irwin~\cite{irwin}. Their general derivation works equally well with
our problem. However our detailed results are different from
theirs. For the sake of completeness, we present their derivation  as
the way applied to our case.

To calculate the metric of the relative moduli space $M^8$, let us
define the tangent vectors of $M^8$. A tangent vector ${\bf Y}=(Y_1,
Y_2, Y_3,Y_4)$ must satisfy the linearized Nahm's equations,
\begin{equation}
\dot{Y}_i+[Y_4, T_i]+[T_4,
Y_i]=\epsilon_{ijk}[T_j, Y_k].
\label{tan1} 
\end{equation}
Since the moduli space $M^8$ is defined by gauge equivalent Nahm
data, the tangent vector should be  orthogonal to 
infinitesimal gauge transformations $\delta T_\mu $ in ${\cal G}_0$, 
\begin{equation}
\sum_{\mu=1}^4 \langle Y_\mu, \delta  T_\mu\rangle =0,
\label{gaugec}
\end{equation}
where the orthogonality is defined with the flat metric on the
infinite dimensional affine space~\cite{nakajima,atiyah}, 
\begin{equation}
ds^2({\bf Y},{\bf Y}') = \sum_\mu \langle Y_\mu,Y_\mu'\rangle
\label{metric1}
\end{equation}
Equation (\ref{gaugec}) can be written in an explicit form,
\begin{equation}
 \dot{Y}_4+\sum_{\mu=1}^4
[T_\mu, Y_\mu]=0.
\label{tan2}
\end{equation}

The procedure of solving Eqs.~(\ref{tan1}) and (\ref{tan2}) for
tangent vectors has been described in Ref.~\cite{dancer1}.  In
general, $Y_\mu$ can be expressed as $Y_\mu= y_{\mu j}(\tau_j/2)$.
Substituting this expression into Eqs.~(\ref{tan1}) and (\ref{tan2}),
we get four closed sets of linear differential equations, whose
nonsingular solutions can be parametrized by eight real parameters
$m_\mu,\; n_\mu$, as follows:
\begin{eqnarray}
& & Y_1=\frac{1}{2}\left[ \dot{f}_1I_1\tau_1+\left(\dot{f}_2I_2
+\frac{m_2}{f_2}\right)\tau_2+\left(\dot{f}_3I_3
+\frac{n_3}{f_3}\right) \tau_3\right], \nonumber\\
& & Y_2=\frac{1}{2}\left[-\dot{f}_1I_2\tau_1 +\left(\dot{f}_2I_1
+\frac{m_1}{f_2}\right)\tau_2 -\left(\dot{f}_3I_4+\frac{n_4}{f_3}
\right) 
\tau_3\right], \nonumber\\
& & Y_3=\frac{1}{2}\left[ -\dot{f}_1 I_3\tau_1 +\left(\dot{f}_2I_4
+\frac{m_4}{f_2}\right)\tau_2 + \left(\dot{f}_3I_1+\frac{n_1}{f_3}
\right)  
\tau_3\right], \nonumber \\
& & Y_4=\frac{1}{2}\left[ \dot{f}_1I_4\tau_1 +\left(\dot{f}_2I_3
+\frac{m_3}{f_2}\right)\tau_2- \left(\dot{f}_3I_2+\frac{n_2}{f_3}
\right) 
\tau_3\right], \label{tan3} 
\end{eqnarray}
where 
\begin{equation}
I_\mu (t)=\int^t_0
dt'\,\left(\frac{m_\mu}{f_2(t')^2}+\frac{n_\mu}{f_3(t')^2} \right).
\label{imu}
\end{equation}
The lower bound of $I_\mu(t)$ is chosen so that they are odd
functions. This makes $Y_\mu$ to satisfy the compatibility condition
$Y_\mu(-t)^T =C Y_\mu(t) C^{-1}$, which is  implied  from
Eq.~(\ref{cond}).  This is the tangent vector on the representative
point  (\ref{data2}) of $SO(3)\times SO(3)$ orbit. 

The metric on the moduli space $M^8$ is induced from the flat metric
(\ref{metric1}) on the infinite dimensional affine algebra. 
With our solutions (\ref{tan3}), the general result is
\begin{equation}
ds^2 ({\bf Y},{\bf Y}')=\sum_{\mu=1}^4 \bigl[(g_1+g_1^2X) m_\mu
m^{\prime}_\mu +(g_2+g_2^2X)n_\mu n^{\prime}_\mu + g_1 g_2 X (m_\mu
n^{\prime}_\mu+n_\mu m^{\prime}_\mu)\bigr],
\label{metric2}
\end{equation}
where 
\begin{eqnarray}
& & X(k,D)= f_1(1)f_2(1)f_3(1), \nonumber\\
& & g_1(k,D)=\int_0^1\frac{dt}{f_2(t)^2}, \nonumber\\
& & g_2(k,D)=\int_0^1\frac{dt}{f_3(t)^2}.
\end{eqnarray}

We can calculate the metric by finding the tangent vector at a generic
point of $M^8$, which can be obtained by the $SO(3)\times SO(3)$
spatial and gauge rotations of Nahm data (\ref{data2}).  Due to the
$SO(3)\times SO(3)$ symmetry of the metric, the general metric can
be found if it is known near the identity.  We want to relate the
coordinates $m_\mu, n_\mu$ of the tangent space at the specific point
to the infinitesimal changes of the parameters $k, D$ and the
infinitesimal $SO(3)\times SO(3)$ transformations~\cite{irwin}. This
corresponds basically the rotation of a rigid body around three
principal axes. Similar to the rigid body case, we can find the
metric once we know the moment of inertia around each principal
axes, which are the coordinate axis for our Nahm data (\ref{data0})
and (\ref{data1})~\cite{dancer1}.  The kinetic part for the rigid body
case is expressed in terms of the left invariant one-forms
\begin{eqnarray}
& & \sigma_1 =- \sin \psi d \theta  + \cos\psi\sin\theta d\varphi,
\nonumber \\ 
& & \sigma_2 = \cos\psi d\theta + \sin\psi\sin\theta d\varphi,
\nonumber \\
& & \sigma_3 = d\psi + \cos d \varphi, 
\end{eqnarray}
which correspond to the infinitesimal spatial rotations around three principal
axes. The corresponding left-invariant one-forms for the gauge
rotations are $\check{\sigma}_i$.  The relations we seek
are
\begin{eqnarray}
& & m_1 = -\frac{1}{2}d\lambda_1, \nonumber \\
& & n_1 = -\frac{1}{2}d\lambda_2, \nonumber\\
& & m_2 =  \lambda_1 \sigma_3, \nonumber\\
& & n_2 =- \frac{g_1\lambda_1}{1+g_2 X}\sigma_3 
   + \frac{1}{\sqrt{g_2+g_2^2X}}\left\{b_3\sigma_3 -c_3
(\frac{f_1(1)}{f_2(1)}\sigma_3 - \check{\sigma}_3) \right\}, \nonumber\\
& & m_3 = \frac{g_2 \lambda_2}{1+g_1 X}\sigma_2 - \frac{1}{\sqrt{g_1
+ g_1^2X}}\left\{ b_2\sigma_2 - c_2 (\frac{f_1(1)}{f_3(1)}\sigma_2 -
\check{\sigma}_2)\right\}, \nonumber\\
& & n_3 = -\lambda_2 \sigma_2, \nonumber\\
& & m_4 = -\frac{g_2(\lambda_1-\lambda_2)}{g_1+g_2} \sigma_1 -
\frac{1}{\sqrt{(g_1+g_2)(1+(g_1+g_2)X)}}
\left\{b_1\sigma_1 - c_1 (\frac{f_3(1)}{f_2(1)}
\sigma_1-\check{\sigma}_1)\right\}, \nonumber\\
& & n_4 = \frac{g_1(\lambda_1-\lambda_2)}{g_1+g_2} \sigma_1
-\frac{1}{\sqrt{(g_1+g_2)(1+(g_1+g_2)X)}} 
\left\{b_1\sigma_1 - c_1 (\frac{f_3(1)}{f_2(2)}
\sigma_1-\check{\sigma}_1)\right\},
\label{mneq}
\end{eqnarray}
where $\lambda_1,\lambda_2$ are given in Eq.~(\ref{lambda12}), 
\begin{eqnarray}
& & b_1 = k^2D^2 \sqrt{\frac{g_1^2}{(g_1+g_2)(1+(g_1+ g_2)
X)}},\nonumber \\
& & b_2 = \frac{g_2 D^2}{\sqrt{g_1 + g_1^2 X}}, \nonumber\\
& & b_3 = \frac{g_1 (1-k^2)D^2 }{\sqrt{g_2 + g_2^2 X}},
\end{eqnarray}
and
\begin{eqnarray}
& & c_1 = f_2(1)f_3(1)\sqrt{\frac{g_1+g_2}{1+(g_1+g_2)X}}, \nonumber \\
& & c_2 = \frac{\sqrt{g_1 + g_1^2X}}{f_2(1)g_1}, \nonumber \\
& & c_3 = \frac{\sqrt{g_2+g_2^2X}}{f_3(1)g_2}.
\end{eqnarray}

Once we replace the parameterization (\ref{mneq}) into the metric
(\ref{metric2}), we would have got the explicit form for the metric of
the moduli space $M^8$. Rather than doing this, let us study the  metric
bit by bit.  The two dimensional space ${\cal Y}^2$ is the
geodesically complete space made of eight copies of the $k-D$ space,
$N^5/Spin(3)$. These eight copies are discussed in the remark after
Eq.~(\ref{data2}).  This space describes the motion of the monopoles
with the vanishing  $SU(2)$ electric charge and zero angular momentum. 
The metric of this space from Eqs.~(\ref{lambda12}), (\ref{metric2})
and (\ref{mneq})   is
\begin{equation}
ds^2_{{\cal Y}^2}=\frac{1}{4}\biggl\{ X(g_1 d\lambda_1 + g_2 d\lambda_2)^2
+g_1d\lambda_1^2+g_2d\lambda_2^2 \biggr\}. 
\end{equation}
Figure 3 shows this space in terms of two coordinates, which in the
shaded region are $D$ and $E\equiv\sqrt{1-k^2}D$. The above metric at
origin is smooth with $D,E$ playing cartesian coordinates.  The origin
corresponds to the spherically symmetric configuration. Two coordinate
axes correspond to two hyperbolic configurations, which are symmetric
along real spatial $x^2,x^3$ coordinate axes.  The diagonal lines do
to one trigonometric one, which is symmetric along real spatial $x^1$
axis.  The boundary curves correspond to the Atiyah-Hitchin
configurations, where the massless monopole have been moved to spatial
infinity.

\begin{center}
\leavevmode
\epsfxsize=2.0in
\epsfbox{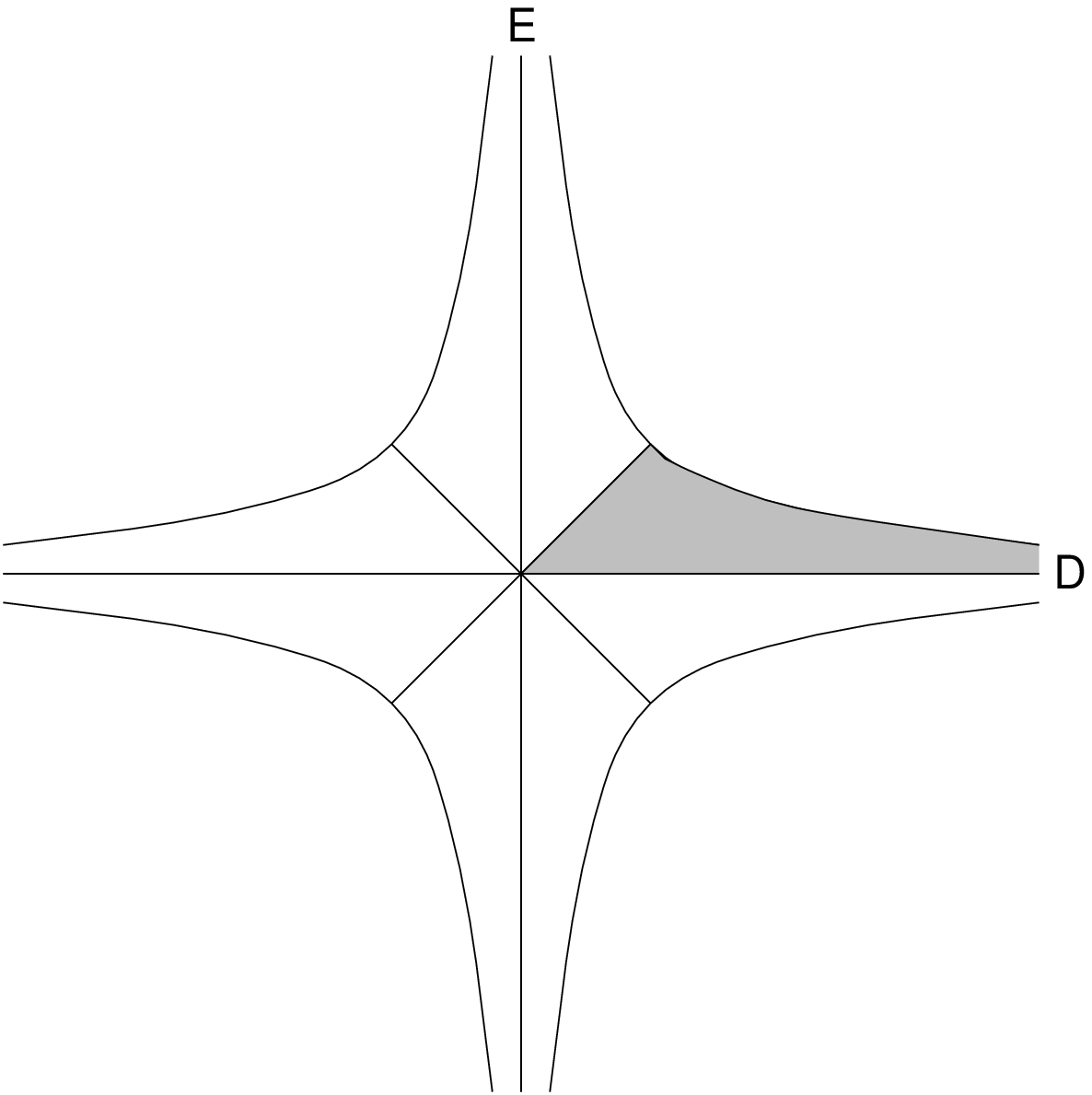}
\end{center}
\begin{quote}
{\bf Figure 3}: A sketch of the geodesically complete space in D-E
coordinates. The shaded region corresponds to the $N^5/Spin(3)$ space.
\end{quote}

\vspace{0.5cm}

While we have not studied in detail the geodesic motion on this space,
one can see from symmetry that the trigonometric solutions with
velocity pointing to the origin will remain trigonometric after the
configuration passes through the origin. With the similar velocity,
the hyperbolic solutions remain hyperbolic, which is consistent with a
picture of noninteracting two monopoles for the hyperbolic case. This
contrasts to  Dancer's case where the trigonometric configurations
changes to the hyperbolic case, and vice versa. The configuration with
 infinite cloud size would remain the Atiyah-Hitchin configurations
and the boundary curve shows the 90 degree scattering of these
monopoles.

The metric on $N^5=M^8/SU(2)$ with the  $Z_2\times Z_2$ isotropic
group  is 
\begin{equation}
ds^2_{N^5} = ds^2_{{\cal Y}^2}
+ a_1\sigma_1^2+a_2\sigma_2^2+a_3\sigma_3^2
\end{equation}
with 
\begin{eqnarray}
& & a_1 = k^4 D^4 \frac{g_1g_2}{g_1+g_2}, \nonumber \\
& & a_2 = D^4 \left\{ g_2+ \frac{g_2^2X}{1+ g_1 X} \right\}, \nonumber \\
& & a_3 = (1-k^2)^2D^4\left\{g_1 + \frac{g_1^2 X}{1+g_2 X}\right\}.
\end{eqnarray}
Here one uses the orthogonality condition for the tangential vectors of
$N^5$ to that of gauge rotation~\cite{dancer1,irwin}, which can be
found from Eq.~(\ref{mneq}) by dropping terms depending on $b_i$ and
$c_i$.  There is no cross term for the invariant one-forms, which is
consistent with $Z_2\times Z_2$ isotropy group of $N^5$.  This metric
describes the monopole dynamics with zero $SU(2)$ electric charge but
 perhaps with nonzero orbital angular momentum. Figure 4 shows two
massive monopoles (two half doughnuts on the $x_3$ axis) with generic
cloud size and three principal axes. In zero cloud size $k=1$, the
metric is symmetric under the rotation around the $x_3$ axis so that
$a_1=a_2$ and $a_3=0$.  In the trigonometric case $k=0$, the metric is
symmetric under the rotation around the $x_1$ axis so that $a_1=0$ and
$a_2=a_3$.

\begin{center}
\leavevmode
\epsfxsize=2.0in
\epsfbox{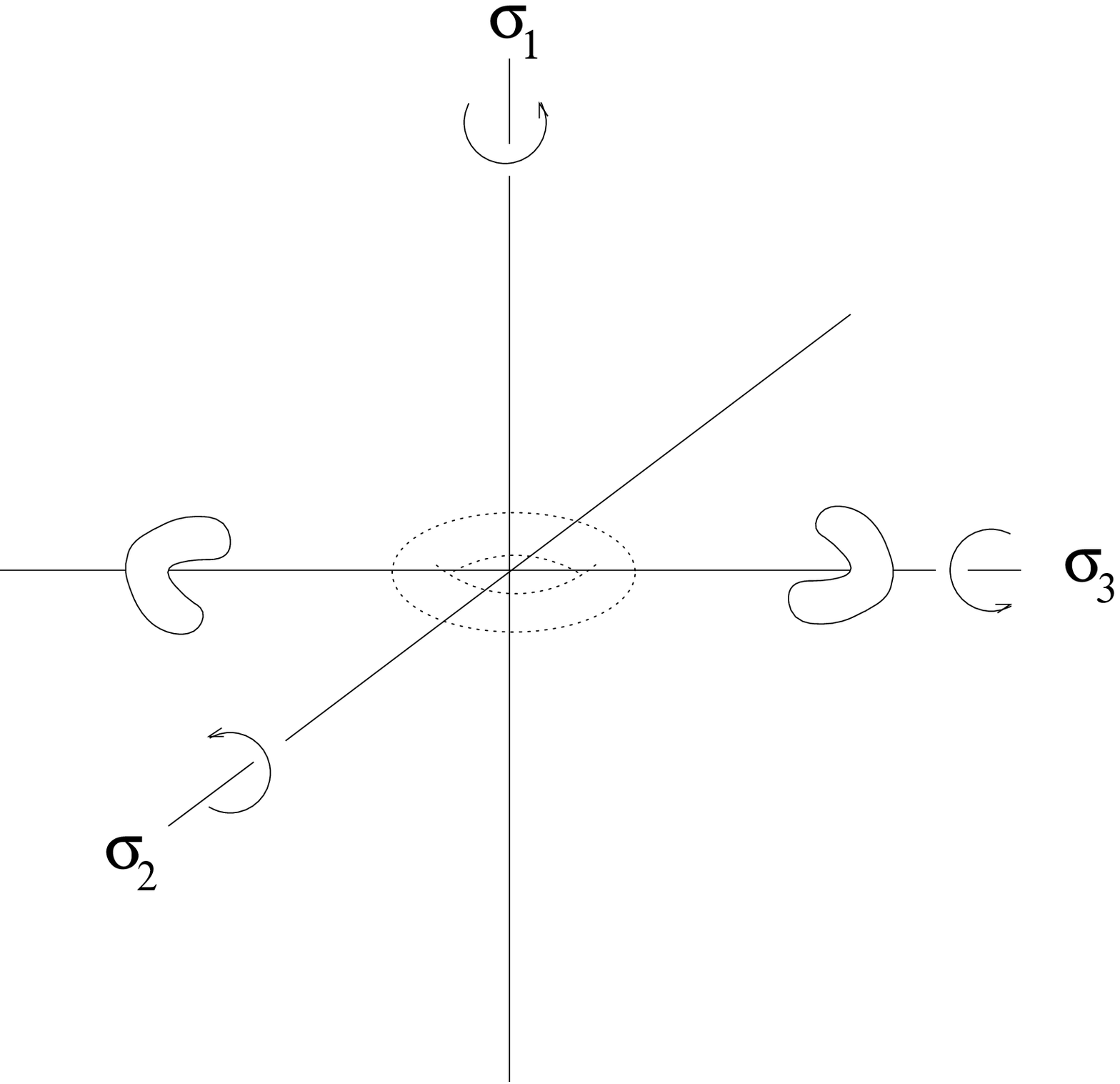}
\end{center}
\begin{quote}
{\bf Figure 4}: The massive monopole with finite size cloud. 
The central doughnut indicates the symmetric axis $x_1$. 
\end{quote}

\vspace{0.5cm}

{}From Eqs.~(\ref{metric2}) and (\ref{mneq}), we  get the full metric on
 $M^8$, which is 
\begin{eqnarray}
ds^2_{M^8} = & & \frac{1}{4}\biggl\{ X(g_1 d\lambda_1 + g_2 d\lambda_2)^2
+g_1d\lambda_1^2+g_2d\lambda_2^2 \biggr\} \nonumber \\
& &  + a_1\sigma_1^2+a_2\sigma_2^2+a_3\sigma_3^2  + \left\{ b_1
\sigma_1 - c_1 \left( \frac{f_3(1)}{f_2(1)}\sigma_1 -
\check{\sigma}_1\right)\right\}^2 \nonumber\\ 
 & &  +\left\{ b_2 \sigma_2 - c_2 \left(
\frac{f_1(1)}{f_3(1)}\sigma_2 - \check{\sigma}_2\right)\right\}^2
+\left\{ b_3 \sigma_3 - c_3 \left(
\frac{f_1(1)}{f_2(1)}\sigma_3 - \check{\sigma}_3\right)\right\}^2.
\label{fullm}
\end{eqnarray}
This metric is hyperk\"ahler. The isometric group is $SO(3)\times
SO(3)$. The $SO(3)$ global gauge transformation is tri-holomorphic and
the $SO(3)$ spatial rotation rotates three complex structures of the
manifold.  There are several interesting limits of this metric.  When
the cloud size is smallest with $k=1$, its Nahm date is the hyperbolic
case (\ref{hyper1}) and the above metric becomes
\begin{equation}
ds^2_{\rm hyper} = dD^2 + D^2 \sigma_1^2+D^2 \sigma_2^2 + D\tanh D
\,\check{\sigma}_1^2 + D\coth D\, \check{\sigma}_2^2 + 
\check{\sigma}_3^2.
\label{hyperm}
\end{equation}
The moments of inertia for internal gauge transformations are nonzero
exactly as we argued in the previous section. Especially when
spherically symmetric case with $D=0$, the coefficient of
$\check{\sigma}_1$ vanishes, and those of $\check{\sigma}_2$ and $
\check{\sigma}_3$ become identical, i,plying  the $S^2$ gauge orbit
space. In large separation $D>>1$, the inertia for $\check{\sigma}_1$
and $\check{\sigma}_2$ become identical. The inertia for
$\check{\sigma}_3$ is constant, which corresponds to $t^3(\alpha)$
dyonic excitations discussed in the next paragraph  after
Eq.~(\ref{hyp}).

There are two  axially symmetric solutions, that is,
hyperbolic and trigonometric. When we include  internal
global gauge rotations which preserve the  axial symmetry, we obtain
two dimensional surfaces of revolution.  The metric for 
the trigonometric case with $k=0, 0<D<\pi/2$  is
\begin{equation}
ds^2_{{\rm trig}^2}=\sec^2D (1+D\tan D)(1+\frac{\sin D \cos D}{D}) \left[
dD^2 + \frac{D^2 (\sigma_1-\check{\sigma}_1)^2}{(1+D\tan D)^2} \right],
\label{revol1}
\end{equation}
where $\sigma_1-\check{\sigma}_1$ can be put into a rotation $d\alpha$
around internal and angular angles.  As $D\rightarrow \pi/2$, the
metric (\ref{revol1}) becomes
\begin{equation}
ds^2=d\rho^2 + \frac{1}{4}\rho^2 d\alpha^2
\end{equation}
where $\rho=2\sqrt{\pi/(\pi - 2D)} $. In this limit the massless
monopole moves out from the localized massive monopoles, and so the
non-Abelian cloud is expected to be more and more spherical with the
flat $R^4$ moduli space as in Ref.~\cite{kleen}.  The above
metric is then a section of $R^4$ with a radial variable $\rho$ as we
will see in a moment. In the physical space, the massless cloud size
is of order $\rho^2$. The non-Abelian component of the gauge field will
change its behavior from $1/r$ to $1/r^2$ as one crosses the this
radius $\rho^2$.  Another axially symmetric case is hyperbolic one
with $k=1, 0<D<\infty$, whose metric is
\begin{equation}
ds^2_{{\rm hyper}^2} = dD^2 + \check{\sigma}_3^2
\label{revol2}
\end{equation}
Clearly this flat metric  is  a part of the metric (\ref{hyperm}).

The limit of large cloud size can be found in the region where $K(k)-D
<< 1$. In the previous section, we argued that Nahm data in this case
becomes that for the Atiyah-Hitchin case.  In this limit one can show
easily the metric (\ref{fullm}) becomes
\begin{eqnarray}
ds^2 = & & d\rho^2
+\frac{\rho^2}{4}\left\{(\sigma_1-\check{\sigma}_1)^2 + 
(\sigma_2 + \check{\sigma}_2)^2 + (\sigma_3 + \check{\sigma}_3)^2 \right\} \\
& & + \frac{b^2}{K^2}dK^2 +a^2\sigma_1^2+b^2\sigma_2^2 +c^2\sigma_3^2 + {\cal
O}(\rho^{-1})
\end{eqnarray}
where $\rho=2\sqrt{D/(K(k)-D)}$, $K=K(k)$ and
\begin{eqnarray}
& & a^2 = \frac{K(K-E)(E-(1-k^2)K)}{E}, \\
& & b^2 = \frac{EK(K-E)}{E-(1-k^2)K}, \\
& & c^2 = \frac{EK(E-(1-k^2)K)}{K-E},
\end{eqnarray}
where $E$ being the second complete elliptic integral $E(k) =
\int_0^{\pi/2}\,d\theta \sqrt{1-k^2\sin^2\theta}.$ This shows that the
asymptotic space is a direct product of $R^4$ and the Atiyah-Hitchin
space. As in Ref.~\cite{kleen}, we expect that the metric of the
massless cloud space approaches that of flat $R^4$, which is exactly
what the above limit shows. A combination of orbital and gauge
angular variables needs to be introduced~\cite{irwin} to make this $R^4$
explicit.

The part of the moduli space metric we can calculate independently
from  Nahm's formalism is the asymptotic metric, which is valid
when the mutual distance between monopoles is  large. This can be done
by studying the interaction between dyons in large
separation~\cite{gibbons,klee2} and taking the massless limit.  In the
center of mass frame, the relative positions between the massive
$\beta^*$ monopoles and the massless $\alpha^*$ monopole are ${\bf
r}_1$ and ${\bf r}_2$ as shown in Fig.~5. The relative position
between two massive monopoles is ${\bf r}={\bf r}_1+{\bf r}_2$.

\begin{center}
\leavevmode
\epsfxsize=2.0in
\epsfbox{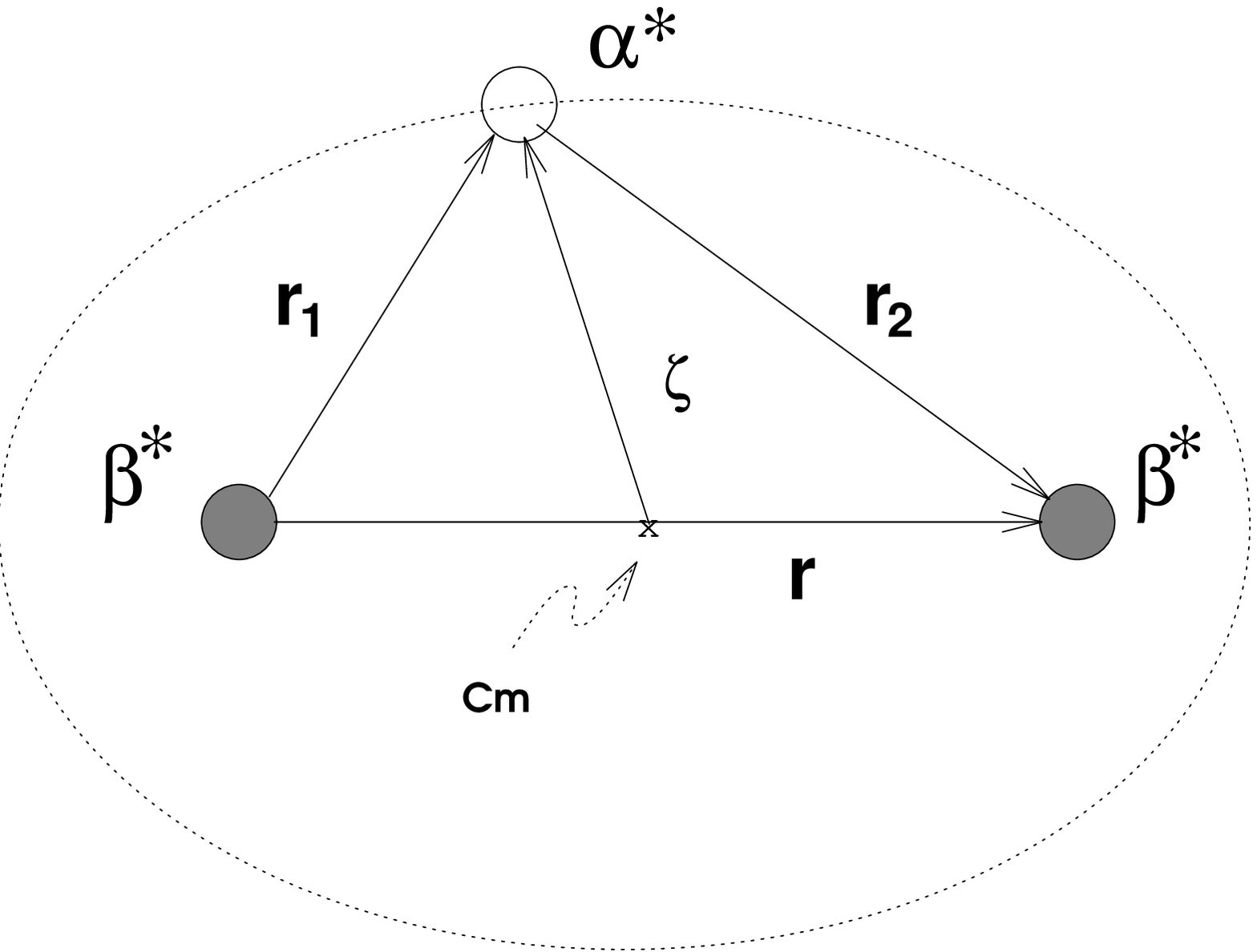}
\end{center}
\begin{quote}
{\bf Figure 5}: The parameters of the asymptotic metric. The center of
mass is at the middle of the line connecting two massive $\beta^*$
monopoles. 
\end{quote}

In terms of the relative positions and the relative angles
$\psi_a,a=1,2$, the asymptotic form of the metric for the relative
moduli space $M^8$  is
\begin{equation}
ds^2=\sum_{a,b}^2 \left [ G_{ab}d{\bf r}_a\cdot d{\bf r}_b
+(G^{-1})_{ab}D\psi_a D\psi_b\right ],
\label{asympt}
\end{equation}
where
\begin{equation}
 G_{ij}=\left( \begin{array}{cc}
                   1+\frac{1}{r_1} -\frac{1}{r} & 1-\frac{1}{r}\\ 
                   1-\frac{1}{r} & 1+\frac{1}{r_2}-\frac{1}{r}
                  \end{array}\right),
\end{equation}
\begin{equation}
D\psi_a=d\psi_a+ {\bf w}({\bf r}_a)\cdot d{\bf r}_a
    -{\bf w}({\bf r})\cdot d{\bf r},
\end{equation}
with the Dirac potential ${\bf w}$ such that $\nabla\times {\bf
w}({\bf r})=\nabla (1/r)$. If we have removed the direct interaction
between two identical massive monopoles, the above metric is identical
to the Taubian-Calabi metric of the $SU(4)\rightarrow U(1)\times SU(2)
\times U(1)$ case. Since the non-Abelian cloud of a massless monopole
is independent of the direct interaction, the $SU(2)$ orbit on
non-Abelian cloud would again be the three-dimensional ellipsoid
defined by $r_1+r_2={\rm constant}$ as shown in Fig.~5.  This fact can
be seen easily by adapting the argument for the $SU(4)$ case in
Ref.~\cite{kleen}.  In the large cloud size limit, one can compare the
exact metric (\ref{fullm}) and the above metric.  We see
$K/(K-D)\approx r_1+r_2$ and $r\approx -\ln \sqrt{1-k^2}$. The
condition $r_1+r_2>> r$ for large size cloud becomes $K(k)-D <<1$.

Now we are in position to learn more about the four dimensional space
$M^4({\mbox{\boldmath $\zeta$}})$ defined by the moment map
(\ref{moment}).  {}For Nahm data~(\ref{data0}) and (\ref{data2}) we
get ${\mbox{\boldmath $\zeta$}}=(\zeta,0,0)$ with
\begin{equation}
\zeta= D\sqrt{1-k^2}\frac{{\rm sn}_k(D)}{{\rm cn}_k(D)}
\end{equation}
The general Nahm data is obtained from that in Eq.~(\ref{data2}) by
spatial and gauge rotations. Thus $\zeta$ would be a function of
rotational and gauge parameters. Now we see that when $\zeta=0$, we
have $k=1$. This corresponds to the hyperbolic data (\ref{hyper1})
with the minimal size of non-Abelian cloud. (In Dancer's case the
hyperbolic data is expected to have the minimum cloud size, and is
different from the $\zeta=0$ case except in large separation.) The
four dimensional metric for this case can be obtained from the metric
(\ref{hyperm}) and is the flat $R^3\times S^1$,
\begin{equation}
ds^2= dD^2 + D^2(\sigma_1^2+\sigma_2^2) +\check{\sigma}_3^2
\end{equation}
Note that the gauge rotation $\check{\sigma}_2$ changes the value of
$\zeta$ as it transforms the hyperbolic data (\ref{hyper1}). In the
language of Ref.~\cite{kleen}, it moves the massless monopole from the
origin. When $\zeta = \infty$, we have $D=K(k)$, which means that
massless monopole has been removed, resulting in the Atiyah-Hitchin
metric.  Thus we see that $M^4({\mbox{\boldmath $\zeta$}})$
interpolates between the $M^4(0)=R^3\times S^1$ and the Atiyah-Hitchin
space $M^4(\infty)$.

In terms of the asymptotic form of the metric (\ref{asympt}), the
$U(1)$ rotation of $SU(2)$ gauge rotates $\psi_1\rightarrow
\psi_1+\epsilon$ and $\psi_2\rightarrow \psi_2-\epsilon$, whose moment
map is
\begin{equation}
{\mbox{\boldmath $\zeta$}}  = \frac{{\bf r}_1-{\bf r}_2}{2},
\end{equation}
as shown in Fig.~5.  As  ${\mbox{\boldmath $\zeta$}}$ increases from zero to
infinity, the size of the non-Abelian cloud increases from zero to
infinity, consistent with the picture discussed in the previous paragraph. 
Also, we can trivially  obtain the asymptotic form of the metric for
$M^4(\zeta)$,
\begin{equation}
ds^2=G\, d{\bf r}^2 + G^{-1}(d\psi + {\bf W}\cdot d{\bf r})^2,
\label{hyperq}
\end{equation}
where 
\begin{equation}
G=1 + \frac{1}{2|{\bf r}+2{\mbox{\boldmath $\zeta$}}|} + \frac{1}{2|{\bf
r}-2{\mbox{\boldmath $\zeta$}}|} - \frac{1}{|{\bf r}|}
\end{equation} 
and ${\bf W}$ is decided from the relation $\nabla G = \nabla \times
{\bf W}$. Clearly this hyperk\"ahler quotient can be done by holding
the position of the massless monopole at $\mbox{\boldmath $\zeta$}$
relative to the center of mass and let massive monopoles to move
around, interacting each other and with the massless
monopole.\footnote{Even in the maximally broken case, ther eis a
conserved $U(1)$ and so the hyperk\"ahler quotient make sense. The
mass parameter of $\alpha^*$ monopole does not lead to any additional
paramter on $M^4({\mbox{\boldmath $\zeta$}})$ since it turns out to
scale the positions ${\bf r}$ and ${\mbox{\boldmath $\zeta$}}$, as far
as the asymptotic metric (\ref{hyperq}) is concerned. This is not a
view shared by Ref.~\cite{houghton}.}  This process breaks not only
the rotational $SO(3)$ symmetry but also the global gauge symmetry
$SO(3)$. We do not think that there is any remaining symmetry on the
$M^4(\zeta)$ for $0<\zeta<\infty$.

\section{Conclusion}

We have studied a purely Abelian BPS monopole configuration made of
two identical massive monopoles and one massless monopole in the
theory where the gauge group $Sp(4)$ is spontaneously broken to
$SU(2)\times U(1)$. We approached this problem by finding the
solutions for the corresponding Nahm's equations under proper boundary
and compatibility conditions. We have used the ADHMN construction to
get the spherically and axially symmetric field configurations, which
are consistent with the field theory picture.  {}From the analysis of
the axially symmetric solutions, we have come to understand the role
of the non-Abelian cloud and its size.  Then the explicit form of the
metric on the eight dimensional moduli space of the relative motion is
found.  By studying the metric in various limit, we see the metric for
the moduli space of the Nahm data is consistent with what is expected
from the monopole dynamics. 

We have also studied the metric of various submanifold of this space.
Our work provide a further support to the idea that the Nahm's
approach for the BPS monopole configurations and their moduli spaces
is valid in general. Our work leads also to some insight on the
characteristics of the non-Abelian cloud and the gauge orbit. It is
interesting to note that the spherically symmetric solution has
nonzero inertia for some of unbroken gauge transformations.

{}From the previous experiences, we now see how in principle one may
find the moduli space metric of two identical massive and one massless
monopoles in the theory with $G_2\rightarrow SU(2)\times U(1)$. To get
this, one may start from the theory with $SO(8)\rightarrow
SU(2)^3\times U(1)$ with two identical massive and three distinct
massless monopoles. If one identifies two massless monopoles, then the
configuration would be that of two massive and two distinct massless
monopoles in the theory with $SO(7)\rightarrow SU(2)^2\times
U(1)$. After further identification of all massless monopoles, one
would get the desired configuration in the theory with $G_2$.

The hyperk\"ahler quotient of the 8-dimensional relative moduli spaces
of these configurations is a four dimensional hyperk\"ahler space.  To
find the $M^4({\mbox{\boldmath $\zeta$}}=0)$, one can consider the
asymptotic form of the metric for two massive monopoles with minimum
size cloud. (We overlap the massless monopole on one of the massive
monopole.)  They are $R^3\times S^1$, Taub-NUT or a double covering of
Atiyah-Hitchin, depending whether they are associated with the gauge
group $Sp(4)$, $G_2$ or $SU(3)$, respectively.  When the cloud size
becomes infinite, all these three four-dimensional spaces
$M^4(\infty)$ approach the Atiyah-Hitchin space.

Another direction to explore is to find the moduli space in the case
when massless monopoles become massive so that there are two identical
massive and one distinct massive monopoles.  We think that the moduli
space in the theory where $Sp(4)\rightarrow U(1)^2$ is simpler than
the similar problem in the theory with $SU(3)\rightarrow U(1)^2$.
Also this moduli space has a role to play in the $N=2$
S-duality~\cite{piljin}.  Finding the moduli space will be a
challenge.  Finally it would be very interesting to find out some
structure of the moduli space of three massive and three massless
monopoles in the theory where $SU(4)\rightarrow SU(3)\times U(1)$. We
know the asymptotic form of the metric and it may be good enough.  As
argued in the introduction, these configurations can be regarded as a
magnetic dual of baryons and would imply new insight on the baryon
structure.

\vspace{2cm}

\centerline{\bf Acknowledgments} 

This work is supported in part by the
U.S. Department of Energy. We thank Soonkeon Nam, Erick Weinberg and
Piljin Yi for useful discussions. K.L. would like to thank Aspen Center
for Physics (1997) and APCTP/PIms summer workshop (1997)

\end{document}